# Effect of Sm-, Gd-co-doping on structural modifications in aluminoborosilicate glasses under β-irradiation

E. Malchukova[a*], B. Boizot [a], G. Petite[a] and D. Ghaleb [b]

[a]CEA/DSM/DRECAM, Laboratoire des Solides Irradiés,UMR 7642, CNRS-CEA-Ecole Polytechnique, Ecole Polytechnique, 91128 Palaiseau Cedex, France.

[b]CEA/DEN/DTCD/SECM, Laboratoire d'études du Comportement à Long Terme, CEA Valrho- Marcoule, BP 171, 30207 Bagnols-sur-Cèze Cedex, France.

**Abstract**

Two series of Sm-, Gd-codoped aluminoborosilicate glasses with different total rare earth content have been studied in order to examine the codoping effect on the structural modifications of β-irradiated glasses. The data obtained by Electron Paramagnetic Resonance spectroscopy indicated that relative amount of $Gd^{3+}$ ions located in network former position reveals non-linear dependence on Sm/Gd ratio. Besides, codoping leads to the evolution of the EPR signal attributed to defects created by irradiation: superhyperfine structure of boron oxygen hole centres EPR line becomes less noticeable and resolved with increase of Gd amount. This fact manifests that $Gd^{3+}$ ions are mainly diluted in vicinity of the boron network. By Raman spectroscopy, we showed that the structural changes induced by the irradiation also reveal non-linear behaviour with Sm/Gd ratio. In fact, the shift of the Si-O-Si bending vibration modes has a clear minimum for the samples containing equal amount of Sm and Gd (50:50) in both series of the investigated glasses. In contrast, for single doped glass there is no influence of dopant's content on Si-O-Si shift (in case of Gd) or its diminution (in case of Sm) occurs which is explained by the reduction process influence. At the same time, no noticeable effect of codoping on $Sm^{3+}$ intensity as well as on $Sm^{2+}$ emission or on Sm reduction process was observed.

*Corresponding author. Eugenia Malchukova : CEA/DSM/DRECAM, Laboratoire des Solides Irradiés, Ecole Polytechnique, 91128 Palaiseau Cedex, France. tel.: 33 1 69 33 38 26; fax: 33 1 69 33 30 22; e-mail: genia@poly.polytechnique.fr

# 1. Introduction

The modern concept of radioactive waste management suggests the confinement of radioactive wastes into the compositions of matrices that possess a chemical, mechanical and radiation resistance. High-level nuclear wastes have been vitrified with the use of, for example, aluminophosphate glasses (in Russia) or borosilicate glasses (in France, England and United States). These nuclear waste glasses are often complex materials containing up to thirty different oxides. In order to predict the long term behaviour under irradiation of these complex glass materials, we have focused our work on simplified glass composition, which reproduces the structure and the behaviour of nuclear wastes glasses. It is known that structural evolution of the nuclear glasses under irradiation ($\alpha$, $\beta$, $\gamma$) can modify the confinement properties [weber et al, 1997]. In order to understand structural changes induced by ionizing radiation, the effect of $\beta$-irradiation on the glass structure has been investigated in a first step on aluminoborosilicate glass compositions with compositions similar to the matrix of the complex nuclear glasses [1-3]. Then we have introduced in these simplified nuclear glass compositions different rare earth (RE) doping ions. Because of the similarity of chemical properties between rare earth (RE) elements and actinides (An), some RE ions are used as surrogates for the actinides during the development of actinide-containing radioactive waste glasses [4-6]. Moreover, a certain amount of RE ions also exists as fission products in some actinide-bearing radioactive wastes [7, 8]. Our previous investigations on single RE-doped aluminoborosilicate glasses showed that ionizing radiation leads to the RE reduction [2, 3, 9]. We analyse also that the structural changes at one integrated dose (defects production , glass densification and polymerization increase) are decreasing as a function of RE doping ion content. This influence of RE doping upon the glass structural changes under irradiation seems to be correlated with the relative stability of the RE different oxidation state For example, the structural changes in Gd- or Nd-doped glasses where the RE reduction processes are either negligible or absent at all are not limited by the RE doping content [2,3]. Thus, our recent studies have been directed on the investigation of interplay between structural evolution and reduction process occurring under irradiation in aluminoborosilicate glass and estimation of some general trend for behaviour under irradiation of single RE-doped glasses. But, taking into account that complex nuclear glasses contain mixture of RE ions as fission products and/or as non-radioactive actinides (An) surrogates, the simulation of RE mixed effect is of a great importance from both scientific and technological points. Meanwhile, to the best of our knowledge, the effect of codoping and codopants concentration on the structural modifications in the glass matrices has not been investigated.

In this paper, initiated by previous results of the luminescence of Sm-doped and structural evolution of Gd-doped borosilicate glass under irradiation [2,9], the results of analysis of codoping effect on the changes in glass structure, RE environment and also on RE reduction process are presented. The choice of $Sm^{3+}$ and $Gd^{3+}$ codoped materials is based on the following criteria. The $Gd^{3+}$ ions are paramagnetic and, as our recent study has showed, can be easily observed by EPR even at room temperature [2]. As for samarium, both $Sm^{3+}$ and $Sm^{2+}$ ions can be detected by means of luminescence spectroscopy since they have a very high efficiency of the emission excited in visible ($Sm^{3+}$) and near IR ($Sm^{2+}$) region [9]. Therefore, codoping seems to be a good way for observation of the changes occurring both in glass structure and in RE ions charge state.

## 2. Experimental

Two glass series were fabricated with two total concentration of the $Sm_2O_3$ and $Gd_2O_3$ mixture – 0.17 and 0.34 mol.% of the total RE oxide quantity. All glasses were prepared by mixing proper amounts of $SiO_2$, $Al_2O_3$, $H_3BO_3$, $Na_2CO_3$, $ZrO_2$ (see table 1). The mixture was crushed to a fine powder and homogeneous mixing was ensured. The homogeneous powder was put in a Pt crucible and was gradually heated to 1500°C in an electric furnace during 14 h in air atmosphere. The glass melt was then poured on a copper plate. In order to favour homogenous distribution of rare earth ions into the glasses, another 2 hours melting at 1500°C was effected and the samples were then grinded and mixed again into a homogeneous state. After quenching, glass samples were annealed at 500°C in order to decrease strain.

All samples were irradiated using 2.5 MeV electrons beam generated by a Van de Graaf accelerator. The temperature during irradiation was maintained around 50 °C by water cooling of the sample holder. Glass thickness was about 0.5 mm in order to reach a homogeneous irradiation in the glass volume. $4 \times 10^9$ Gy irradiation dose was reached on each sample with 14 µA electron beam.

Electron Paramagnetic Resonance (EPR) spectra were obtained using a X band ($\nu \approx 9.420$ GHz) EMX Bruker spectrometer at room temperature with a 100 kHz field modulation, 1 G amplitude modulation and 1 mW microwave power. All EPR spectra have been normalized to the samples weight of 100 mg. Error estimation has been done for all data obtained taking into account contribution of the random and systematic errors. Systematic error included uncertainty in the mass measuring. Mass measurement uncertainty was of 0.1 mg. In order to minimize the random error measurements of the spectra were repeated 10 times. All uncertainties have been considered and displayed in the table and on the data in the figures. The microwave frequency has been determined using a Hewlett Packard 5352B frequency counter.

Luminescence and Raman spectra were collected after irradiation on a Labram HR microspectrometer using the 514.5 nm line of an $Ar^+$ laser. Experiments were carried out through x100 Olympus objective and a laser power about 80 mW on the sample was used to avoid significant heating of the samples. All Raman spectra of pristine and β-irradiated glass samples have been normalized to the intensity of the band near 460 cm$^{-1}$.

## 3. Results

### 3.1. EPR spectra

The EPR spectra for pristine and heavily irradiated codoped glass, illustrated in figure 1, show well-known features at g ~ 2.0; 2.8 and 6.0 of "U" spectrum [2, 10], typical for Gd$^{3+}$ ion in vitreous materials as well as in disordered polycrystalline matrices. According to literature, the prominent EPR signal at g ~ 4.8 is also observed in disordered materials [2, 11]. These two groups of features associated with a distinct type of crystal field exists to be known as corresponding to Gd$^{3+}$ ions located in the network modifier (Gd[n.m.]$^{3+}$) and network former (Gd[n.f.]$^{3+}$) positions, respectively [12] (figure 1). One can see that β-irradiation at doses higher than 10$^9$ Gy affects the content of Gd$^{3+}$ ions in network former sites and leads to the appearance of the additional EPR defects line around g ~2.0. We investigated correlation between the EPR lines attributed to the Gd[n.m.]$^{3+}$ and Gd[n.f.]$^{3+}$ sites before and after irradiation for both series Sm-,Gd-codoped aluminoborosilicate glasses (1SG and 2SG) in dependence on Sm/Gd molar ratio that are shown in figure 2a. The obtained ratio between Gd[n.f.]$^{3+}$ and Gd[n.m.]$^{3+}$ sites follows a non-linear behaviour with Sm/Gd ratio accompanied by abrupt increase of Gd[n.f.]$^{3+}$ sites with increase of Gd fraction in the total molar RE content. The tendency is more important for irradiated samples (figure 2a). By contrast, in single Gd-doped glasses, the relative proportion of Gd$^{3+}$ ions located in network former is decreasing with the gadolinium oxide concentration for both pristine and irradiated aluminoborosilicate glasses as it is shown in figure 2b. Sm/Gd Codoping therefore influences the speciation of Gd$^{3+}$ between a former and modifier network positions in both pristine and irradiated glasses.

Figure 3 presents the defects EPR spectra produced by the irradiation for 1SG and 2SG glass samples. We observed that superhyperfine structure of BOHC defects [13] appears with increase of the Sm/Gd ratio. This structure of defect line is noticeable at rather low content of $Sm_2O_3$ in codoped samples for 1SG series (1SGb sample) and becomes significant for single Sm-doped glass (1SGs sample) (figure 3a). However, the ability to discern the structure of the defects diminishes with increase of the total RE content from 0.17(1SG) to 0.34 (2SG) mol.% and resolution can be distinguished only for the samples with Sm/Gd ratio above 3/1 (2SGc) (figure 3b). It is interesting to notice that amount of defects varies with codopant concentration as it shown in figure 4 for both glass series. The defect concentration follows a non-linear dependence on Sm-,Gd-codoping (figure 4) for the lowest irradiation dose which is not similar to the single-doped glass studied before. For all single-doped glasses recently investigated we obtained a decrease of the defect amount with dopant concentration [2, 9]. Therefore, codoping can strongly affect also the defect creation processes at least for the lowest irradiation dose.

*3.2. Raman spectra*

Fig. 5 displays the Raman spectra of irradiated and non-irradiated Sm-,Gd-codoped aluminoborosilicate glasses. One can notice that the band attributed to the Si-O-Si bending vibration modes (at 460 cm$^{-1}$) is shifted in the glasses irradiated with 10$^9$ Gy. The decrease of $Q^2$ species ($Q^n$ species correspond to $SiO_4$ units with *n* bridging oxygens) stretching band that appears around 1000 cm$^{-1}$ in comparison with $Q^3$ ones (1100 cm$^{-1}$) as it can be seen from the figure is also typical for these glasses both without dopants and RE-doped and has been observed previously [2, 9, 14, 15]. But strong overlapping of the intense Sm luminescence arising from 1100 cm$^{-1}$ with Raman peaks of $Q^2$ and $Q^3$ species prevents quantitative estimation of $Q^3/Q^2$ ratio changes as well as observation of other modifications in Raman spectrum. It is possible however to measure the Si-O-Si bending vibration modes shift with the relative fraction of gadolinium oxide (460 cm$^{-1}$) (Fig.6a) which apparently revealed non-linear character. On th contrary, the same dependence of Si-O-Si shift on RE concentration in single Sm- or Gd-doped aluminoborosilicate glasses showed completely different behaviour as it is shown in figure 6b.

## 4. Discussion

Recently it was shown that structural modifications of aluminoborosilicate glasses matrix doped with rare earth ions under the ionizing radiation are correlated with the content and the nature of the RE dopant [2, 3, 15]. It was established that the matrix changes under ionizing radiation like the decrease of average Si-O-Si angle, the glass polymerization increase and also point defect creation processes can be linked to the relative stability of the RE ion redox states. By contrary, for RE ions with negligible reduction processes under ionizing radiation like Gd and Nd [, ], the host glass matrix changes under irradiation doesn't depends on the RE doping content. Double-doped and single doped aluminoborosilicate glasses containing Sm and/or Gd show different behavior under irradiation, evidenced from both EPR and Raman spectroscopy. It could be suggested that mixing of these two dopants in investigated glasses should show monotonic dependence of structural changes and defect creation on RE content in the mixed series from Gd to Sm related with increase of Sm/Gd ratio. In fact as it can be seen from the figure 2a for each glass series (1SG and 2SG) the Gd[n.f.]/Gd[n.m.] follows a non-linear dependence on the Gd content. It is interesting to notice that without mixing the same dependence reveals clear monotonic decreasing character (figure 2 b). Apparently it should be explained by some RE codoping effect.

The EPR lineshape belonging to point defects is changed in both series of glasses with variations in the proportions of Sm, Gd up to single-doped samples. Indeed, we can observe on figures 3a and 3b a broadening of the BOHC superhyperfine structure with the increase of the relative proportion of Gd inside the glass. It is well known that dipole-dipole interaction results in the broadening and non-resolved EPR lines. Thus we can make an assumption that Gd ions are located in the vicinity to BOHC defects created by irradiation. Such assumption of specific distribution of $Gd^{3+}$ ions in borosilicate glasses has been already proposed by Hong Li et al from Raman spectroscopy measurements on borosilicate glasses doped with the highest $Gd_2O_3$ content [16]. But, these BOHC EPR lines changes did not show a Sm/Gd codoping effect. By contrary, it should be noted that the defect concentration produced during irradiation depends on the Sm/Gd relative proportion and reveals clearlty a non-linear character. Moreover, one can see that this effect has a tendency to decrease with increase of the irradiation dose (Fig. 4). One explanation could be that the environment of both Gd and Sm ions (coordination number and redox state) changes also as a function of the dose and therefore could have an influence on the defects production.

N. Ollier et al has observed that defect concentration produced during irradiation in Na/Li and Na/K mixed alkali aluminoborosilicate glasses follows a non-linear behaviour [17-18]. It has be shown also that mixed-alkali effect (MAE) leads to the decrease of alkaline migration under ionizing radiation and therefore to a strong decrease of matrix structural changes under irradiation : namely densification and polymerization increase at one integrated dose. Unlike MAE, RE codoping does not prevent structural changes at integrated dose higher than $10^9$ Gy as it can be seen by Raman spectroscopy (figure 5). The shift of Si-O-Si bending vibration modes around 480 $cm^{-1}$ also revealed clearly a non-linear character with Sm/Gd ratio (figure 6a). On the contrary, for single RE doped irradiated glasses, this ionizing radiation effect is decreasing with the RE content with a slope directly correlated to RE ions reduction efficiency (figure 6b). For example, the increase of Gd amount inside the glass has no effect on the structural modifications observed by Raman spectroscopy (figure 6b). On the other hand, presence of strong reduction process like in case of Sm-doped glasses ($Sm^{3+} + \beta \rightarrow Sm^{2+}$ + defects) limits the structural modifications (figure 6b) and the defects produced at one integrated dose [15]. Therefore, impacts of RE codoping should reveal linear dependence of both defect amount and structural changes as a function of RE content moving along the series from single Gd to single Sm doping case. But, observed correlation between non-linear characters of both structural evolution and defect creation let us suppose that some additional processes exist connected with possible energy transfer between RE ions [20]. Moreover, taking into account that processes accompanying absorption of ionizing radiation by codoped glasses may be different [19], it is obvious that the explanation of RE codoped effect is not so unequivocal and request further investigation with use of luminescence measurements. Indeed, codoping effect of certain rare-earth elements with other ones is well-known and can lead either to enhancement or quenching of the intensity and the duration of RE luminescence [20, 21].

## 5. Conclusion

Our results on structural evolution under β-irradiation of two glass series joins attest the presence of rare earth mixed effect in aluminoborosilicate glasses and its influence on the glass behaviour under irradiation. First, we showed a non-linear character in the speciation of the $Gd^{3+}$ ions residing in the network former sites with the Sm/Gd ratio. The Raman shift of Si-O-Si bending vibration modes induced by β-irradiation in Sm-,Gd-codoped glasses revealed the same non-linear character along with RE concentration as a function of the RE mixing. Finally, the point defect produced during ionizing radiation at one integrated dose follows also a non linear behavior with the relative proportion of Sm and Gd inside the glass. Therefore, it looks apparent that both the glass structure and the structural modifications in β-irradiated glass are affected by the Sm/Gd-codoping. Besides, changes of EPR lineshape attributed to the BOHC defects with addition of Sm ions confirmed that $Gd^{3+}$ ions are

. The present work further enriches the efforts being made in the development of reliable model for interplay of reduction mechanisms and structural changes taking place under irradiation in rare earth doped aluminoborosilicate glasses.

**Acknowledgements**

We are grateful to Thierry Pouthier and Stephane Esnouf for their contribution during external β irradiation experiments. We also thank Gilles Montagnac and Bruno Reynard (Laboratoire des Sciences de la Terre, ENS Lyon, France) for assistance in Raman experiments.

**References**

[1] W.J. Weber, R.C. Ewing, C.A. Angell, G.W. Arnold, A.N. Cormack, J.M. Delaye, D.L. Griscom, L.W. Hobbs, A. Navrotsky, D.L. Price, A.M. Stoneham and M.C. Weinberg, J. Mat. Res. 12 (8) (1997) 1946.

[1] A.Abbas, Y.Serruys, D.Ghaleb, J.M. Delaye, B.Boizot, B.Reynard, G.Calas. Nucl. Instr. and Meth. B 166-167 (2000) 445.

[2] E. Malchukova, B. Boizot, D. Ghaleb, G. Petite, J. of Non_Cryst. Solids 352 (2006) 297.

[3] E. Malchukova, B. Boizot, G. Petite, D. Ghaleb, phys. Stat. sol. (c) 4(3) (2007) 1280.

[4] P. Loiseau, D. Caurant, N. Baffier, L. Mazerolles, C. Fillet, J. of Nuclear Materials 335 (2004) 14.

[5] C. Lopez, X. Deschanels, J.M. Bart, J.M. Nounals, C. Den Auwer, E.Simoni, J. of Nuclear Materials 312 (2003) 76..

[6] F. Thevenet, G. Panczer, P. Jollivet, B. Champagnon, J. of Non_Cryst. Solids 351 (2005) 673.

[7] A.M. Bevilacqua, N.B. Messi de Bernasconi, D.O. Russo, M.A. Audero, M.E. Sterba, A.D. Herredia, J. Nucl.Mater. 229 (1996) 187.

[8] Y.I. Matyunin, A.V. Demin, E.G. Teterin, Glass Phys. Chem. 21 (6) (1995) 432.

[9] E. Malchukova, B. Boizot, G. Petite, D. Ghaleb, J. of Lumin. 111 (2005) 53.

[10] C. M. Brodbeck, L. I. Iton, J. Chem. Phys. 83(9) (1985) 4285.

[11] S. Simon, I. Ardelean, S. Filip, I. Bratu, I. Cosma, Sol. State Com. 116 (2000) 83.

[12] E. Culea, A. Pop, I. Cosma, J. of Magnetism 157/158 (1996) 163-164.

[13] B. Boizot, G. Petite, D. Ghaleb, G. Calas Nucl. Instr. Methods Phys. Res. B 141 (1998). 580; B. Boizot, G. Petite, D. Ghaleb,, G Calas J. Non-Cryst. Solids 283(1-3) (2001) 179.

[15] E. Malchukova, B. Boizot, G. Petite, D. Ghaleb, J. Non-Cryst. Solids 353(24-25) (2007) 2397-2402.

[16] H. Li, Y. Su, L. Li, M. Quian, D.M. Strachan, J. Non-Cryst. Solids 292 (2001) 167.

[17] N. Ollier, B. Boizot, B. Reynard, D. Ghaleb, G. Petite, Nucl. Instr. Methods Phys. Res B 218 (2004) 176.

[18] N. Ollier, T. Charpentier, B. Boizot, G. Petite, J. Phys.: Condens. Matter 16 (2004) 7625.

[19] A.J. Wojtowicz, M. Balcerzyk, E. Berman, A.Lempicki, Phys. Rev. B 49 (1994) 14880-95.

[20] M.J. Elejalde, R. Balda, J. Fernandez, E. Macho, J.L. Adam. Phys. Rev. B 46(9) (1992) 5169.

[21] M. F. Hazenkamp, G. Blasse, Chem. Mater. 2 (1990)105.

Figure Captions

Figure 1. Evolution of EPR spectra of 0.34 mol.% Sm-, Gd-codoped aluminoborosilicate glasses with increase of Sm content.

Figure 2a and 2b. Relative amount of $Gd^{3+}$ ions located in the network former position as a function of Gd2O3 concentration for codoped (a) and single doped (b) glasses- line is the guide for the eyes only.

Figure 3a and 3b. Changes in EPR spectra of (a) 0.17 mol.% (b) 0.34 mol.% Sm-, Gd-codoped aluminoborosilicate glasses (defect region at g ~ 2) with increase of Sm content.

Figure 4. Defect concentration as a function of Sm content for 105 and 109 Gy - line is the guide for the eyes only.

Figure 5. Raman spectra of 0.34 mol.% Sm-,Gd-codoped aluminoborosilicate glasses for different Gd content.

Figure 6a and 6b. ΔRaman shift of Si-O-Si bending vibration modes for (a) 0.17 and 0.34 mol.% Sm-,Gd-codoped aluminoborosilicate glasses as a function of Gd content, (b) Sm- or Gd-doped aluminoborosilicate glasses as a function of Sm or Gd content - line is the guide for the eyes only.

Table 1

|  | 1SGs (1SGg) | 1SGa | 1SGb | 1SGc | 2SGs (2SGg) | 2SGa | 2SGb | 2SGc |
|---|---|---|---|---|---|---|---|---|
| x | 1.00 | 0.25 | 0.50 | 0.75 | | | | |
| y | | | | | 1.00 | 0.25 | 0.50 | 0.75 |
| $SiO_2$ | 59.77 | | | | 59.67 | | | |
| $Al_2O_3$ | 3.84 | | | | 3.83 | | | |
| $B_2O_3$ | 22.41 | | | | 22.38 | | | |
| $Na_2O$ | 12.12 | | | | 12.09 | | | |
| $ZrO_2$ | 1.70 | | | | 1.69 | | | |
| $Sm_2O_3$ ($Gd_2O_3$) | 0.17 | 0.045 (0.126) | 0.087 (0.084) | 0.131 (0.042) | 0.34 | 0.084 (0.253) | 0.174 (0.167) | 0.263 (0.081) |

Table 1: Nominal compositions of 5 oxide aluminoborosilicate glasses (in mol.%)

The first line indicates glass names beginning by 1SG for 0.17 mol. % of mixture $Sm_2O_3+Gd_2O_3$ series and by 2SG for 0.34 mol. % of mixture of $Sm_2O_3+Gd_2O_3$ series. Second and third lines correspond to molar ratio: x, y = $Sm_2O_3/(Sm_2O_3+Gd_2O_3)$ where $Sm_2O_3+Gd_2O_3$ is for 0.17 and 0.34 mol.%, respectively. The systematic error was less than 0.005 mol. % of each component used for glass preparation.

Figure 1

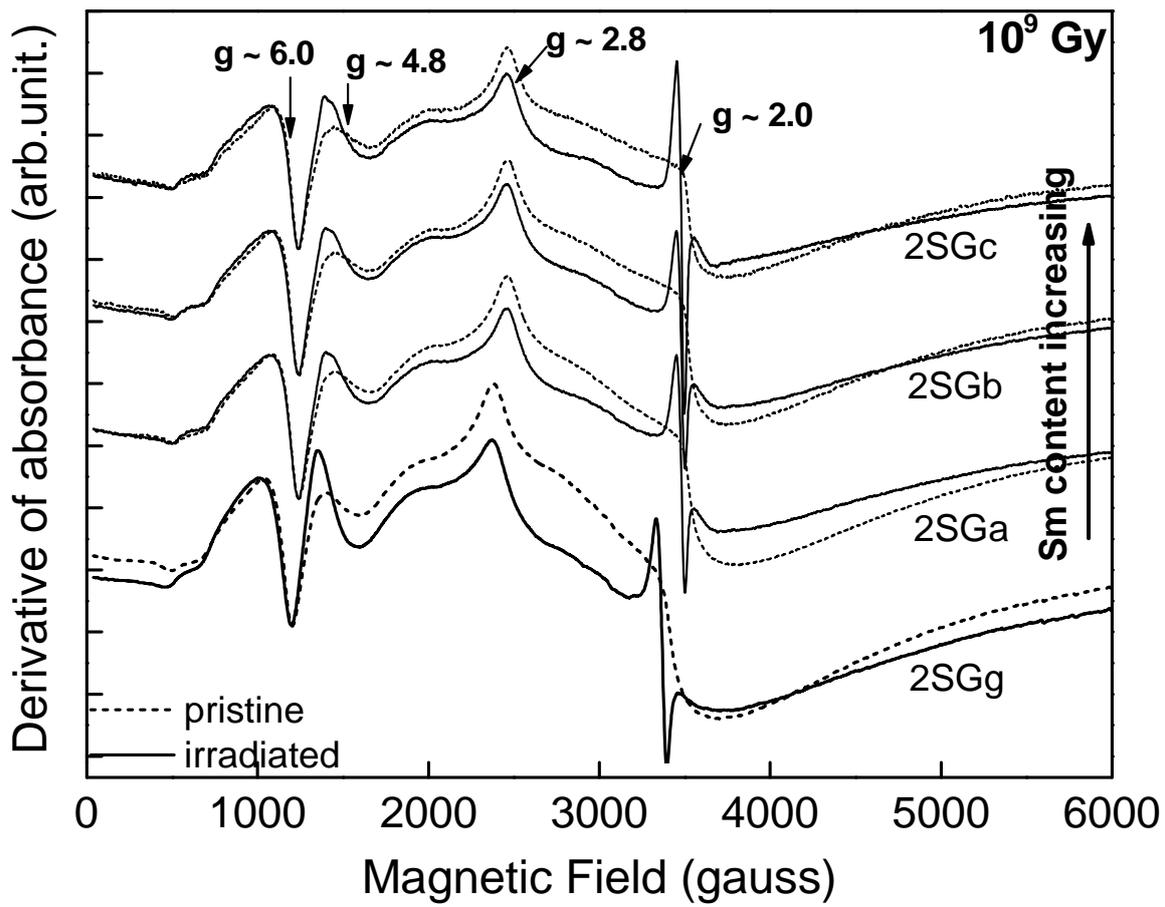

Figure 2a

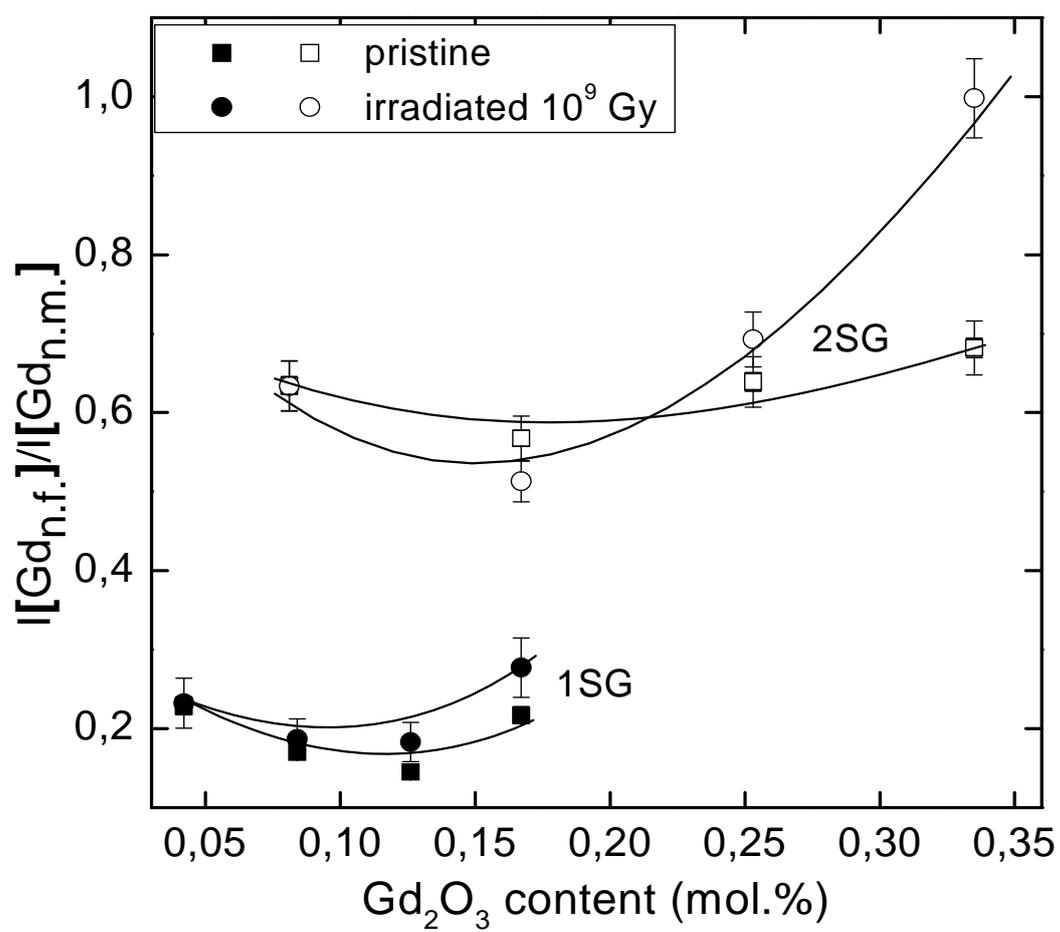

Figure 2b

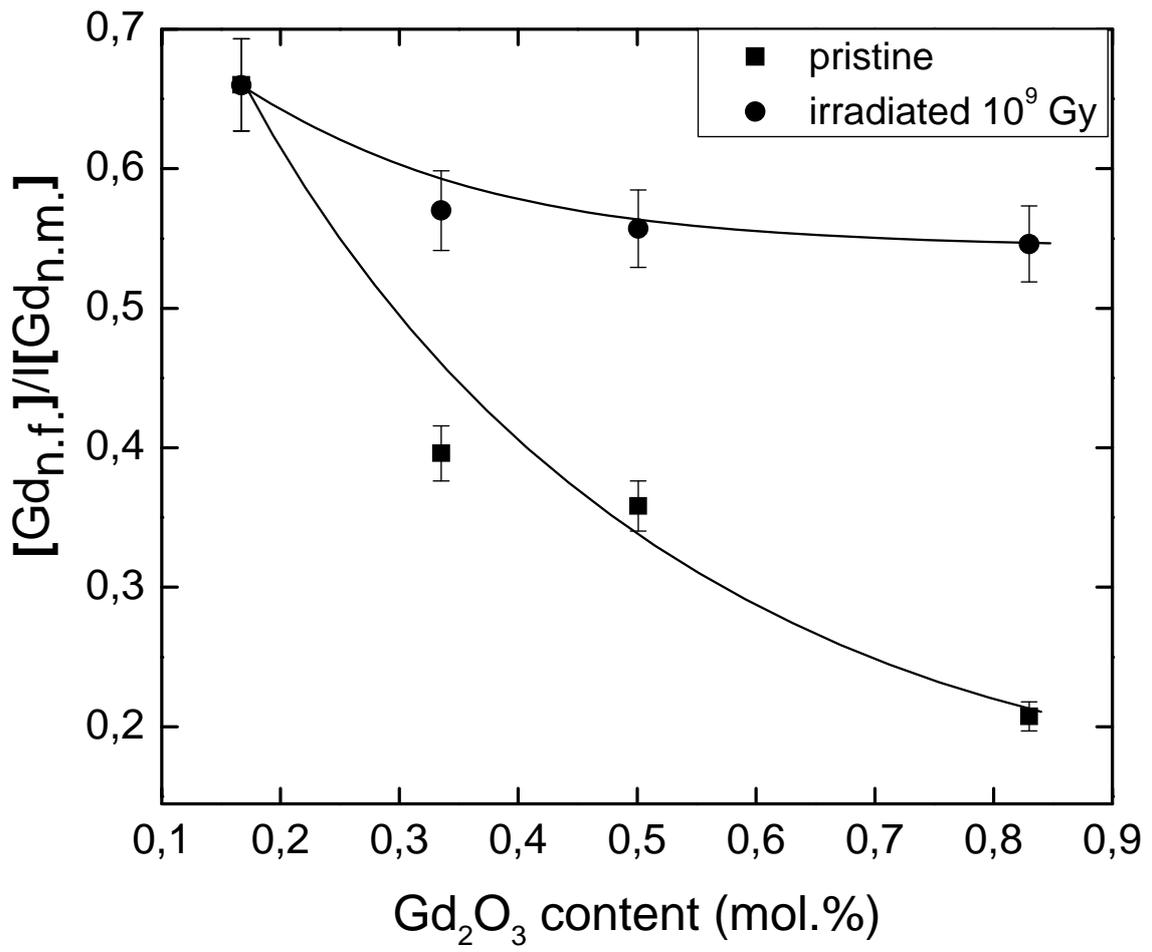

Figure 3a

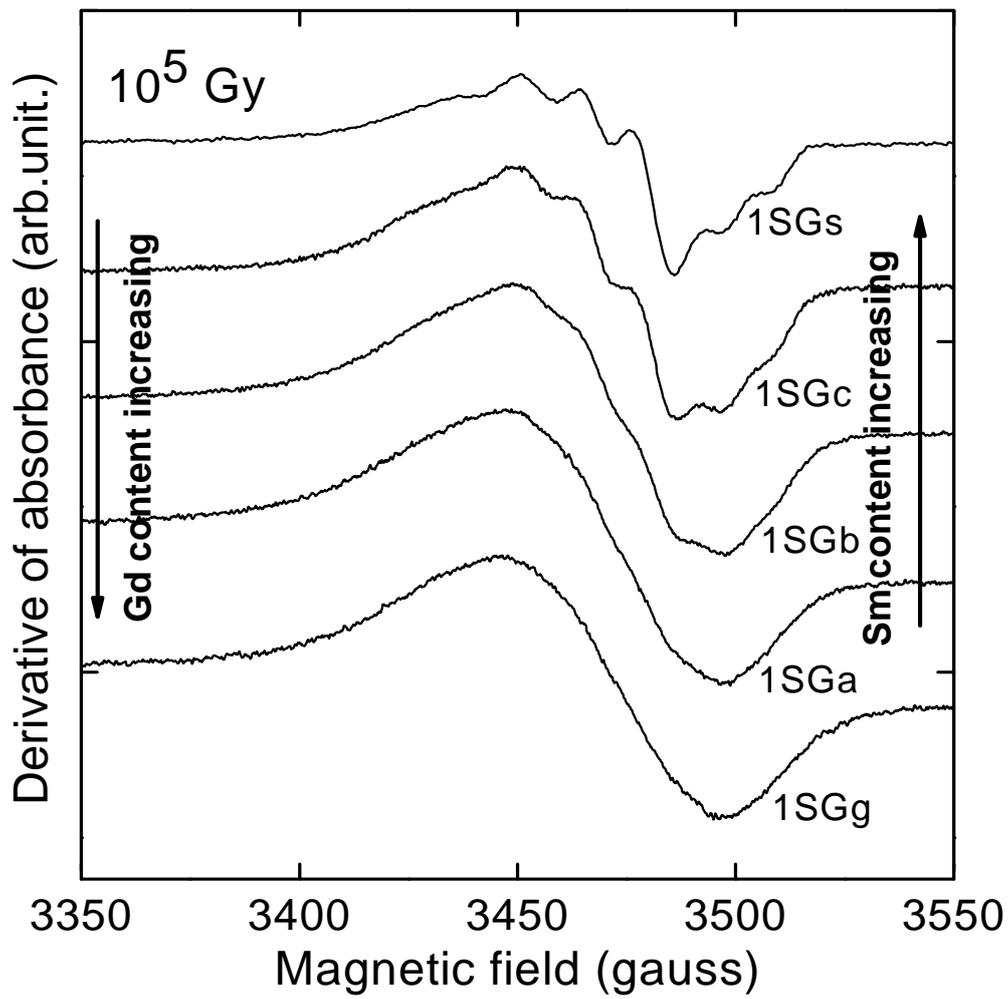

Figure 3b

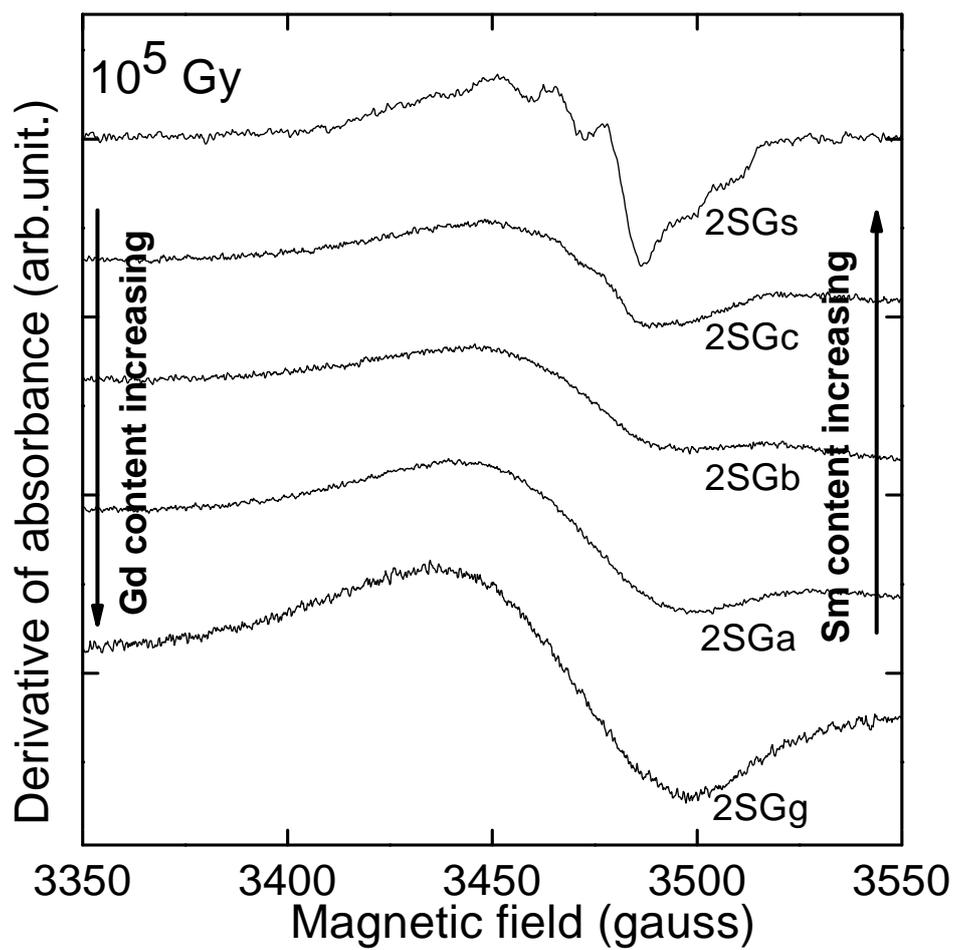

Figure 4

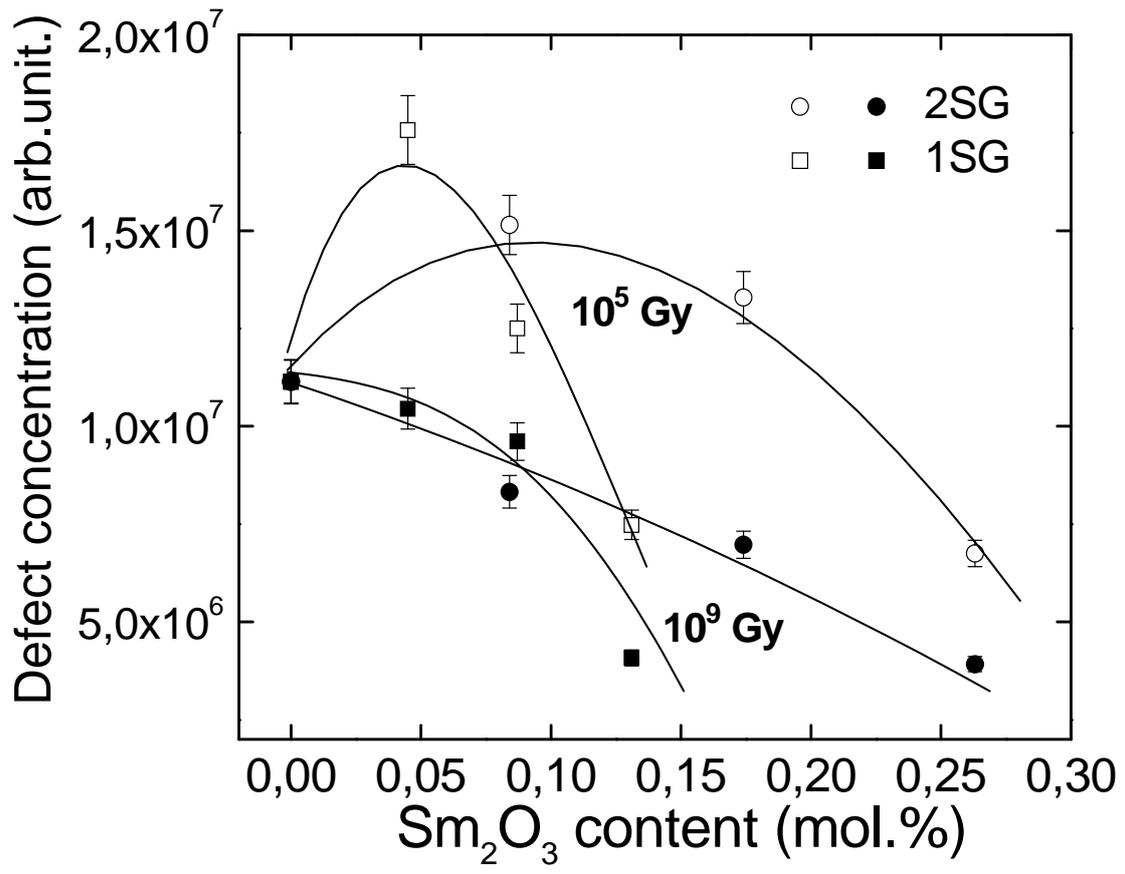

Figure 5

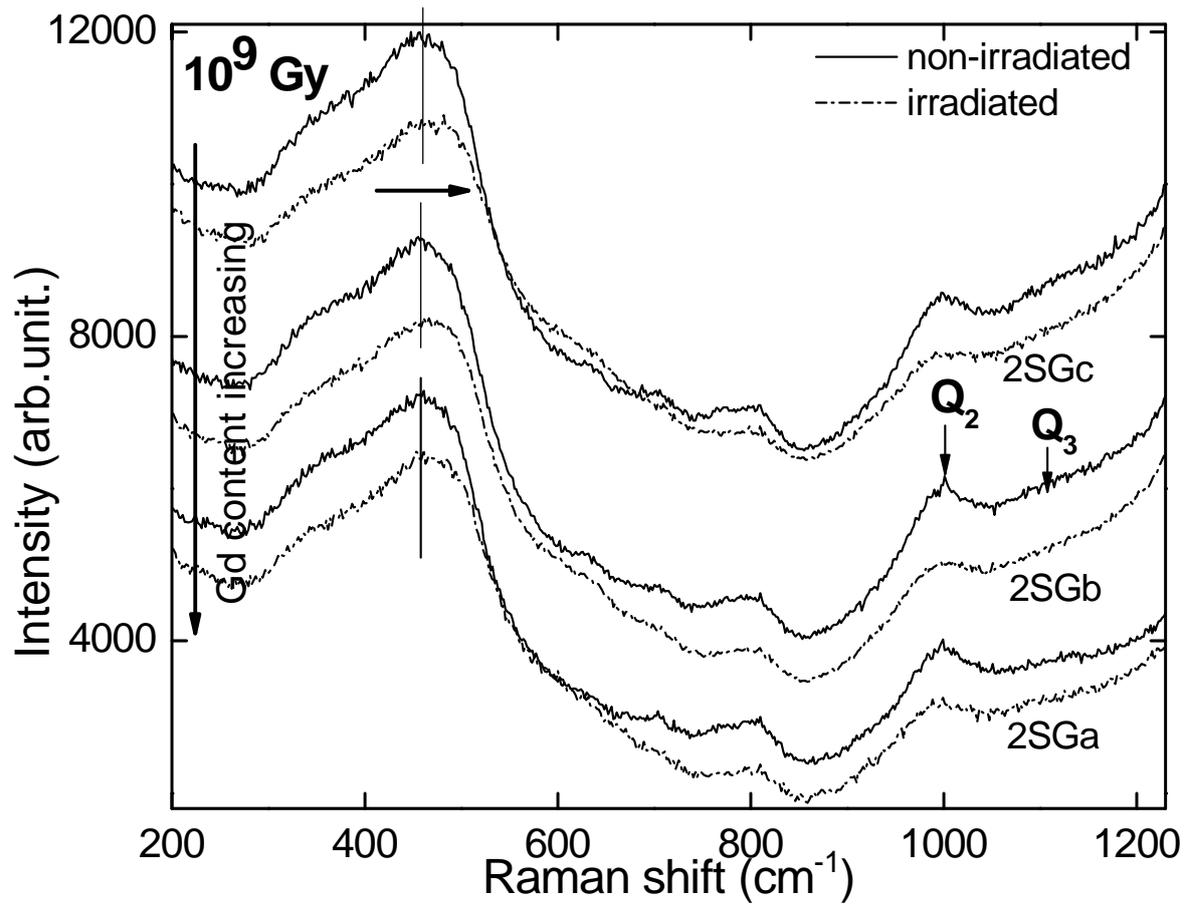

Figure 6a

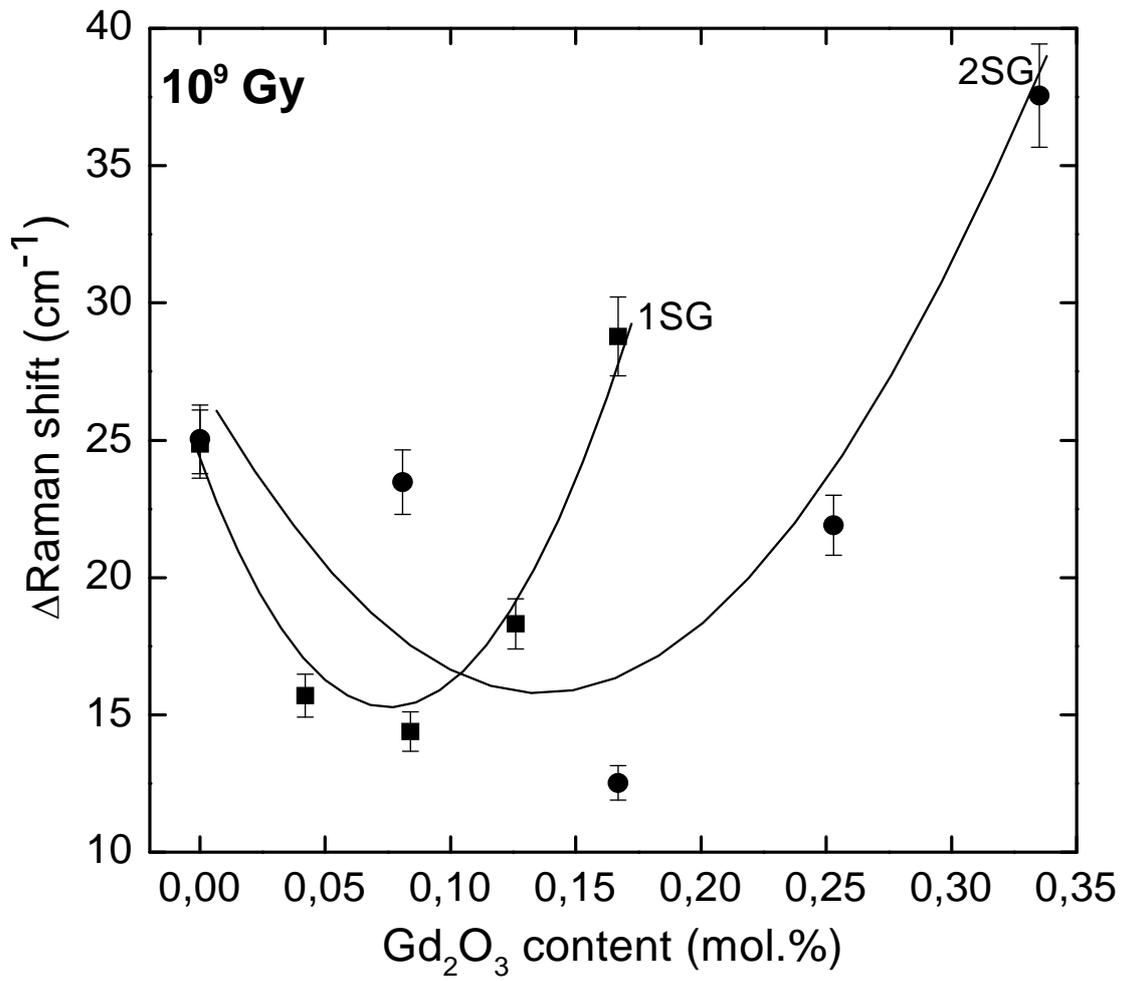

Figure 6b

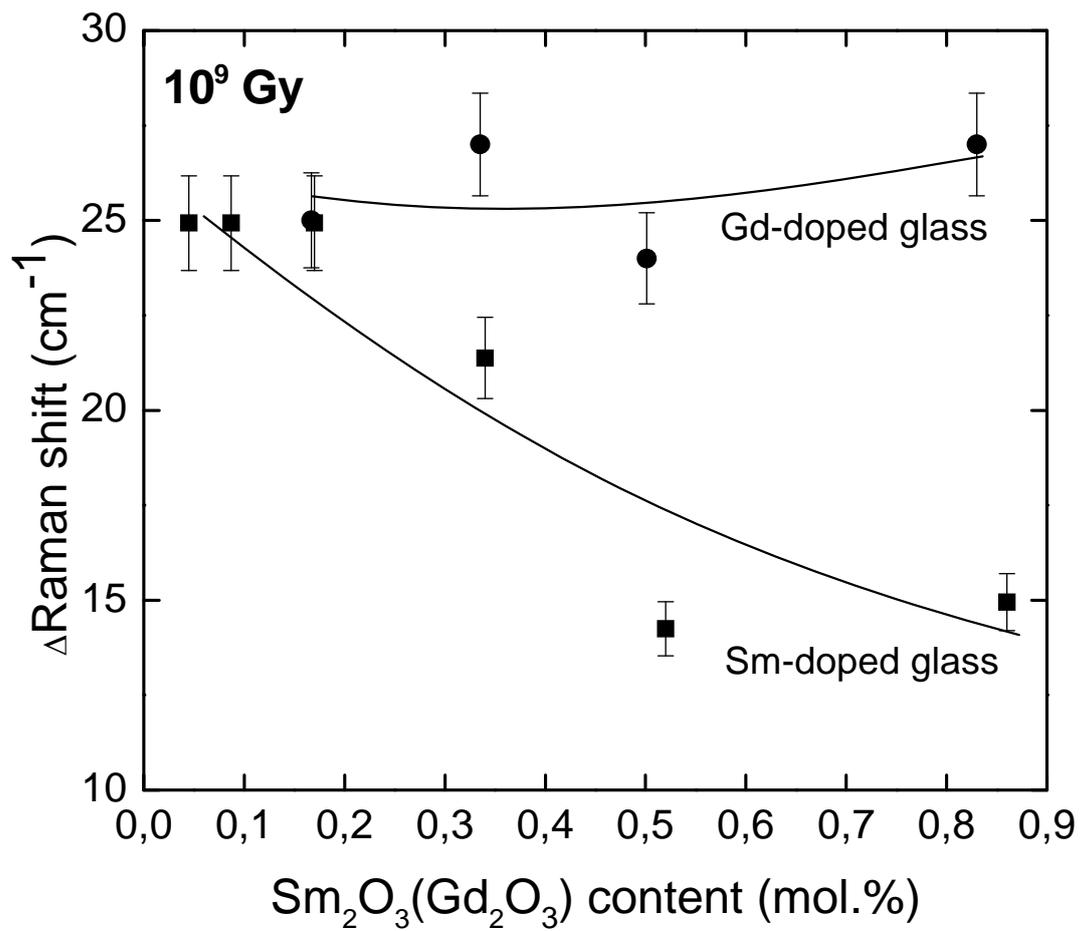